\documentclass[tex,twocolumn,epjc3]{svjour3}
\RequirePackage[T1]{fontenc}
\RequirePackage{graphicx}
\RequirePackage{mathptmx}      
\RequirePackage[numbers,sort&compress]{natbib}
\RequirePackage[colorlinks,citecolor=blue,urlcolor=blue,linkcolor=blue]{hyperref}
\usepackage{amsmath, amssymb}
\usepackage{graphics,bm}
\usepackage{graphicx}
\usepackage{bbold}
\usepackage{slashed}
\usepackage{feynmf}
\usepackage{physics}
\usepackage{wrapfig}
\usepackage{hyperref}
\usepackage[toc,page]{appendix}
\newcommand{\be}{\begin{equation}}
\newcommand{\ee}{\end{equation}}

\journalname{Eur. Phys. J.}
\begin{document}
\title{Finite Density Two Color Chiral Perturbation Theory Revisited}
\author{Prabal Adhikari\thanksref{e1,addr1} \and Soma B. Beleznay\thanksref{e2,addr2}\and Massimo Mannarelli\thanksref{e3,addr3}}
\thankstext{e1}{e-mail: adhika1@stolaf.edu}
\thankstext{e2}{e-mail: belezn1@stolaf.edu} 
\thankstext{e3}{e-mail: massimo@lngs.infn.it}   
\institute{Physics Department, Faculty of Natural Sciences and Mathematics, St. Olaf College, 1520 St. Olaf Avenue, Northfield, MN 55057, United States\label{addr1}
          \and
          St. Olaf College, 1520 St. Olaf Avenue, Northfield, MN 55057, United States\label{addr2}
          \and
          Laboratori Nazionali del Gran Sasso, Via G. Acitelli, 22, I-67100 Assergi (AQ), Italy\label{addr3}
}
%
\date{Received: date / Accepted: date}
\date{\today}
\maketitle
\begin{abstract}
We revisit two-color, two-flavor chiral perturbation theory at finite isospin and baryon density.  We investigate the phase diagram obtained  varying the isospin  and  the baryon chemical potentials,  focusing on the phase transition  occurring when the two chemical potentials are equal  and  exceed  the pion mass (which is degenerate with the diquark mass). In this case, there is a   change in the order parameter of the theory that does not lend itself to the standard picture of first order transitions. We explore this phase transition both within a Ginzburg-Landau framework valid in a limited parameter space and then  by inspecting the full chiral Lagrangian in all the accessible parameter space. Across the phase transition between the two broken phases the order parameter becomes an $SU(2)$ doublet,  with the ground state fixing the expectation value of the sum of the magnitude squared of the pion and the diquark fields. Furthermore, we  find that the Lagrangian at equal chemical potentials is invariant under global $SU(2)$ transformations and construct the effective Lagrangian of the three Goldstone degrees of freedom by integrating out the radial fluctuations. 
\end{abstract}

\section{Introduction}
The study of Quantum Chromodynamics (QCD) is hindered by its inherent non-perturbative nature: the analytical study of QCD is generally not possible except in certain limited regimes. Similarly hindered is the lattice study of QCD in particular at finite baryon density due to the fermion sign problem, see~\cite{Barbour:1986jf, Barbour:1997ej}. On the other hand, lattice QCD studies at vanishing baryonic density and nonzero isospin density are possible for not too large isospin chemical potentials, see~\cite{Alford:1998sd,Kogut:2002zg,Detmold:2012wc,Detmold:2008yn,Cea:2012ev,Endrodi:2014lja, Brandt:2017oyy}. Moreover,  isospin asymmetric matter at nonvanishing magnetic fields cannot be studied using lattice QCD except in a very limited regime~\cite{Endrodi:2014lja, Adhikari:2015wva}. As such QCD practitioners have focused their attention on a wide range of variations of QCD including 't Hooft's large $N_{c}$ limit, see ~\cite{Manohar:1998xv} for a review and~\cite{Kogut:1999iv,Kogut:2000ek} for more recent studies with adjoint quarks and two-color QCD with fundamental quarks~\cite{Kogut:1999iv,Kogut:2000ek}. The primary interest in two-color QCD stems from the fact that at finite baryon density, the fermion determinant is real and positive-definite for two flavors of quarks and as such lattice QCD studies are possible~\cite{Kogut:2001na,Hands:2000ei}. This allows for instance the investigation of color superconductivity due to the formation of diquark condensates and the competition with the chiral condensate. However, it is worth noting that while the fermion determinant remains real and positive definite at finite isospin density with the baryon chemical potential being zero and vice-versa, when both isospin and baryon chemical potentials are simultaneously nonzero, the fermion determinant for the two-flavor case ceases to be positive definite, while remaining real. As such lattice calculations are not possible in this scenario. If both the baryonic  and the isospin chemical potentials are nonzero, the lattice studies require that at least four fermion flavors are present, indeed in this case the fermion determinant is positive definite. The alternative is to study finite density QCD using models such as the Nambu-Jona-Lasinio (NJL) model~\cite{Ratti:2004ra, Andersen:2010vu} and the quark-meson model~\cite{Andersen:2014xxa, Adhikari:2016eef}, which qualitatively reproduce many features of QCD. While unsystematic, these models are useful tools to investigate chiral symmetry breaking, color (de)confinement and exotic phases such as chiral density waves~\cite{Buballa:2014tba}.

Effective field theories also serve as useful tools to study the properties of QCD but, unlike models, the results are systematic with observables computed by  well-controlled approximations~\cite{Weinberg:1978kz}, see also \cite{Pich:1998xt, Holstein:2000ap}. In the present paper we will use chiral perturbation theory ($\chi$PT), which is an effective field theory able to describe many low-energy properties of  QCD~\cite{Scherer:2005ri} by an effective Lagrangian  derived from the global symmetries of QCD. 
The $\chi$PT Lagrangian can be used to systematically reproduce the strong interactions between hadrons by a momentum expansion. At each order in the momentum expansion the global symmetries of QCD fix the form of the various Lagrangian pieces but  the pre-factors, the so-called low energy constants (LECs),  must be determined by different means. The leading order (LO) $\chi$PT Lagrangian depends on only two LECs: the pion decay constant, $f_{\pi}$, and the pion mass, $m_{\pi}$, which are both known at high precision. Remarkably, the  LO$\chi$PT  is sufficient to accurately describe the phase structure of QCD at   $\mu_I \sim m_\pi$~\cite{Son:2000xc,Kogut:2001id,Carignano:2016lxe}, possibly including magnetic fields and  finite temperature effects~\cite{Loewe:2002tw,Loewe:2004mu,Loewe:2016wsk}. The field content of the $\chi$PT Lagrangian (similar to other effective theories) depends on the phenomena and on the energy scale of interest; in the present work we focus on a  realization  that includes only the low-lying pionic and diquark  states of two-color two-flavor systems.


The focus of this paper will indeed be two-color, two-flavor QCD at finite baryonic and isospin densities. In particular we study the nature of the  phase transitions occurring by varying the isospin and the baryonic chemical potentials. Of particular interest is the  case of equal chemical potentials exceeding the pion mass. Previous studies indicate that in this case the phase transition is of the first order~\cite{Splittorff:2000mm}. This result relies on a qualitative argument and on a quantitative analysis. The qualitative argument is that at the phase transition  two different condensates compete and this naturally leads to a first order phase transition. The quantitative argument relies on the inspection of the baryonic and isospin number densities, which indeed are discontinuous at the phase transition  occurring when $\mu_B=\mu_I>m_\pi$.  We reanalyze   this phase transition, finding that this is not a standard first order phase transition. First we consider a  Ginzburg-Landau (GL) expansion valid for second order  and weak first order phase transitions. This analysis shows that for characterizing the phase transition at $\mu_B=\mu_I \gtrsim m_\pi$ it  suffices to consider an expansion including terms up to quadratic order in the fields. In contrast a first order phase transition would require the inclusion of terms of sixth order in the fields. Moreover, the GL expansion naturally leads to a Gross-Pitaevskii (GP) Lagrangian,  see \cite{Carignano:2016lxe} for a GP expansion for  three-color QCD. The GP Lagrangian  is formally the same obtained in ultracold mixture of Bose gases, but with the important difference that in our case all the parameters depend on the chemical potentials and that the number densities of the two species are not fixed. Moreoveor, at the phase transition the intra-particle coupling becomes equal to the inter-particle interaction, and that all the single particle parameters are the same, leading to an enhanced symmetry group. Then, we turn to the analysis of the full Lagrangian, and notice that with an appropriate parameterization of the fields  it is possible to explicitly show a  flat direction of the potential. The associated mode becomes massless for  $\mu_B=\mu_I \gtrsim m_\pi$, and indeed the system has an enlarged $SU(2)$ symmetry.  This means that the phase transition is not a standard first order phase transition because not all the derivatives of the potential with respect to the fields are discontinuous.


The paper is organized as follows: After a a brief review of two-color chiral perturbation in  Sect.~\ref{sec:review}, we present  a first  discussion of the phase diagram of two-flavor, two color $\chi$PT. In Sect.~(\ref{sec:phase}) we investigate in more detail the nature of the various phase transitions. Finally, in Section~(\ref{sec:effective}), we construct the low energy effective Lagrangian by integrating out the radial mode in the $SU(2)$ invariant theory valid when the isospin chemical potential is equal to the baryon chemical potential. In the Appendix we obtain some useful thermodynamical relations.

\section{Review of two-color chiral perturbation theory}
\label{sec:review}
\subsection{Lagrangian}
We begin with a brief review of two-color, two-flavor QCD. In the chiral limit and at zero chemical potentials, two-color QCD possesses an expanded Pauli-G\"{u}rsey symmetry with the symmetry group $SU(2N_{f})$, with $N_{f}=2$. The vacuum breaks the symmetry down to the symplectic group $Sp(4)$ resulting in a Goldstone manifold $SU(4)/Sp(4)$, which according to Goldstone's theorem has five Goldstone modes: these are identified with two oppositely charged pions, a neutral pion and two oppositely charged baryons (diquarks). The $SU(4)/Sp(4)$, coset space can be mapped to a $SO(6)/SO(5)$ coset space, see  \cite{Smilga:1994tb, Brauner:2006dv},  with all the soft fields  collected in a unimodular matrix $\Sigma$   parametrized as  follows
\be
\Sigma= \sum_{k=1}^{6} n_k \tilde\Sigma_k\,,
\ee
where $ \sum_{k=1}^{6} n_k n_k =1$  and $\tilde\Sigma_k$ is a set of independent antisymmetric $4\times 4$ matrices. For $\Sigma$ to be unitary the basis matrix has to be properly chosen, see  \cite{Brauner:2006dv}.
We use the same  non-linear parametrization,  but with $n_0=\cos\rho$ and $n_i=\hat\phi_i\sin\rho $ for $i=1,\dots,5$. The
$\rho$ field is the ``radial coordinate" and $\hat{\phi}_i$ are unimodular  fields, $ \sum_{i=1}^{5}  \hat{\phi}_{i}\hat{\phi}_{i}=1$. In more detail, the proposed nonlinear realization is  
\begin{equation}\label{eq:representation}
\Sigma=\cos\rho\ \Sigma_{0}+i\sin\rho\ \Sigma_{i} \hat{\phi}_{i}\ ,
\end{equation}
where   the basis matrices are
\begin{equation}
\begin{split}
\Sigma_{0}&=
\begin{pmatrix}
  0 & -\mathbb{1} \\
  \mathbb{1} & 0
\end{pmatrix},\ 
\Sigma_{1}=
\begin{pmatrix}
  -i\tau_{2} & 0 \\
  0 & -i\tau_{2}
\end{pmatrix},\\ 
\Sigma_{2}&=
\begin{pmatrix}
  0 & \tau_{1} \\
  -\tau_{1} & 0
\end{pmatrix},\ 
\Sigma_{3}=
\begin{pmatrix}
  \tau_{2} & 0 \\
  0 & -\tau_{2}
\end{pmatrix},\\ 
\Sigma_{4}&=
\begin{pmatrix}
  0 & \tau_{2} \\
  \tau_{2} & 0
\end{pmatrix},\ 
\Sigma_{5}=
\begin{pmatrix}
  0 & \tau_{3} \\
  -\tau_{3} & 0
\end{pmatrix},
\end{split}
\end{equation}
where $\mathbb{1}$ is the $2$$\times$$2$ identity matrix and $\tau_{a}$, with $a=1,2,3$, are the  Pauli matrices.
This basis is related to that of  \cite{Brauner:2006dv} as follows: $\Sigma_0=\tilde\Sigma_1$ and $\Sigma_i=-i \tilde\Sigma_{i+1}$. 

As we will see, the choice of the representation in Eq.~\eqref{eq:representation} has the merit of clarifying the properties of the ground state in a way that is maybe even more transparent than in \cite{Brauner:2006dv}. The relation with diquark and pion fields can be easily obtained defining   the  real scalar fields $\phi_i = \rho \hat{\phi}_{i}$, and then  noting that
\be
\label{eq:N_pi} N=\frac{\phi_{1}+i\phi_{3}}{\sqrt{2}} \,\,\,\text{and}\,\,\,  \pi =\frac{\phi_{2}+i\phi_{4}}{\sqrt{2}}\,,
\ee
correspond to the diquark  and charged pion fields, respectively. The neutral pion can be identified with the $\phi_5$ field.

Nonvanishing baryon and isospin densities can be accounted for by the introduction of  covariant derivatives~\cite{Gasser:1983yg,Kogut:2000ek,Kogut:1999iv,Splittorff:2000mm,Brauner:2006dv}. The   $\mathcal{O}(p^{2})$ Lagrangian turns out to be
\begin{equation}
\mathcal{L}_{2}=\frac{f_\pi^2}{4}\Tr(D_{\mu} \Sigma D^{\mu} \Sigma^{\dagger})+\frac{f_\pi^2 m_\pi^2}{2} \Tr(\Sigma_{0}^{\dagger} \Sigma)\ ,
\end{equation}
where  $f_\pi$ is the pion decay constant  and  the pion mass  is degenerate with the diquark mass, with $f_\pi^2 m_\pi^2 = (m_u +m_d) G$ by the Gell-Mann-Oakes-Renner relation, and we have assumed that quarks only have Dirac masses.  We note that the diquark and  pion masses are degenerate even if one considers a nonvanishing mass difference between up and down quarks. This is due to the fact that both diquarks and pion fields are  up-down quarks (or antiquarks)   states and that the their masses depend linearly on the quark masses.

Finally, given our convention for the $\Sigma_i$ basis, and noting  that $\Sigma_0$ anti-commutes with the baryon number and with the third component of isospin,  the covariant derivatives are defined as 
\begin{align}
D_{\mu}\Sigma &=\partial_{\mu}\Sigma-i\delta_{\mu 0}\{\mu_{\rm B}B+\mu_{I}I_3,\ \Sigma \}\,,\nonumber\\
D_{\mu}\Sigma^{\dagger}&=\partial_{\mu}\Sigma^{\dagger}+i\delta_{\mu 0}\{\mu_{\rm B}B+\mu_{I}I_3,\ \Sigma^{\dagger} \}\ ,
\end{align}
with  $B={\rm diag}\left(\frac{1}{2},\frac{1}{2},-\frac{1}{2},-\frac{1}{2}\right)$ the baryon matrix  and  $I_3={\rm diag}\left (\frac{1}{2},-\frac{1}{2},-\frac{1}{2},\frac{1}{2} \right )$ the matrix corresponding to the third component of isospin. Note that the only LECs of the $\mathcal{O}(p^{2})$ Lagrangian are the pion decay constant and the pion mass: the introduction of the isospin and baryonic chemical potential does not require any extra LECs.

\subsection{Phase diagram}
The homogeneous ground state of the system can be extracted by minimizing the classical potential, which using the representation in Eq.~\eqref{eq:representation} has the form
\begin{equation}
\label{eq:potential}
\begin{split}
V=&-f_{\pi}^{2}m_{\pi}^2 \Big [2(\cos\rho-1)\Big.\\
&+\left.\sin^{2}\rho\left \{\gamma_{B}^{2}\left (\hat{\phi}_{1}^{2}+\hat{\phi}_{3}^{2} \right )+\gamma_{I}^{2}\left (\hat{\phi}_{2}^{2}+\hat{\phi}_{4}^{2} \right ) \right \}\right ]\ ,
\end{split}
\end{equation}
where 
\begin{equation}\label{eq:gammas}
\gamma_{I}\equiv\frac{\mu_{I}}{m_{\pi}} \,\,\,\text{and}\,\,\, \gamma_{B}\equiv\frac{\mu_{B}}{m_{\pi}}\ .
\end{equation} 
Since the potential is at a minimum when the fields $\hat{\phi}_{i}$, with $i=1,\dots,4$, are at a maximum, we take 
\be\label{eq:phi5} \langle \hat{\phi}_{5} \rangle=0\,,\ee 
meaning that the condensate never points in this direction. Since this direction corresponds to the $\pi_0$ field, it follows that the neutral pion does not play any role in characterizing the phase transitions. We distinguish four different regimes: 
\begin{enumerate}
\item \label{item:case1} If both $\gamma_{B}<1$ and $\gamma_{I}<1$ the normal vacuum with
 \be \langle \rho\rangle=0\ee 
 is favored. 
\item \label{item:case2} If  $\gamma_{B}>\gamma_{I}$ and $\gamma_{B}\ge1$ the ground state is obtained maximizing the $\hat{\phi}_{1}^{2}+\hat{\phi}_{3}^{2}$ term in the potential, therefore
\begin{align}
\langle \hat{\phi}_{1}^{2}+\hat{\phi}_{3}^{2} \rangle&=1\\
\langle \hat{\phi}_{2}^{2}+\hat{\phi}_{4}^{2} \rangle&=0\\
 \langle \rho  \rangle=\rho_N  &=\arccos\left(\frac{1}{\gamma_{B}^{2}} \right )\,,
\end{align}
meaning that the radial field lies in the  $(\phi_1,\phi_3)$ plane. This condition can be expressed in terms of the diquark and pion fields  \be\label{eq:cond1} \langle | N| \rangle = \frac{ \rho_N  }{\sqrt{2}}\,\,\,\,\,\,  \text{and} \,\,\,\,\,\, \langle | \pi| \rangle =0\,.\ee
\item\label{item:case3} At variance, if $\gamma_{I}>\gamma_{B}$ and $\gamma_{I}\ge1$ the ground state is obtained maximizing the $\hat{\phi}_{2}^{2}+\hat{\phi}_{4}^{2}$ term in the potential, leading to 
\begin{align}
\langle \hat{\phi}_{1}^{2}+\hat{\phi}_{3}^{2}\rangle&=0\\
\langle \hat{\phi}_{2}^{2}+\hat{\phi}_{4}^{2}\rangle &=1\\
 \langle \rho  \rangle=\rho_\pi &=\arccos\left(\frac{1}{\gamma_{I}^{2}} \right )\,,
\end{align}
which can equivalently be expressed by
\be\label{eq:cond2} \langle | \pi| \rangle = \frac{\rho_\pi }{\sqrt{2}}\,\,\,\,\,\,  \text{and} \,\,\,\,\,\,\langle | N| \rangle =0\,.\ee
\item\label{item:case4} Finally if $\gamma_{B}=\gamma_{I}=\gamma\ge1$, the ground state is characterized by
\begin{align}
\langle \hat{\phi}_{1}^{2}+\hat{\phi}_{2}^{2}+\hat{\phi}_{3}^{2}+\hat{\phi}_{4}^{2} \rangle&=1\\
 \langle \rho  \rangle&=\arccos\left(\frac{1}{\gamma^{2}} \right )\,,
\end{align}
where the first equality is automatically satisfied because of Eq.~\eqref{eq:phi5}.
In this case \be\label{eq:vevtot} \left \langle \sqrt{| N|^2 + | \pi|^2}  \right\rangle = \frac{ \langle \rho  \rangle}{\sqrt{2}} \,,\ee meaning that the radial coordinate can pick up any direction in the $(\phi_1,\phi_2,\phi_3,\phi_4)$ space. 
\end{enumerate}

\begin{figure}
  \includegraphics[width=0.45\textwidth]{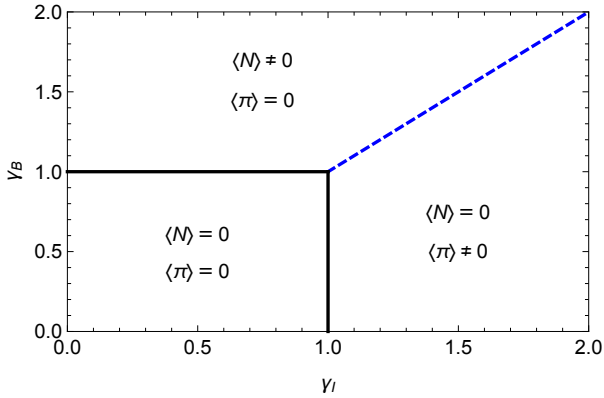}
\caption{Phase diagram as a function of the isospin and baryonic chemical potentials, see Eq.~\eqref{eq:gammas}. Solid lines correspond to second order phase transitions between the normal phase and the broken phases; the dashed (blue) line corresponds to the  phase transition between the diquark and the pion condensed phases. }
\label{fig:phase_diagram}       
\end{figure}

The corresponding phase diagram is reported in Fig.~\ref{fig:phase_diagram} and is equivalent to the one reported in \cite{Splittorff:2000mm}. The transition along the solid lines is second order, indeed $\langle \rho \rangle=0$ along this phase transition lines. On the other hand for $\gamma_{I}=\gamma_{B}>1$ there is a peculiar phase transition. One may expect it to be first order, because the two broken phases are characterized by different condensates. However, along the dashed line  the two condensates do  not separately acquire a vev, but only their combination, given in Eq.~\eqref{eq:vevtot}, does. The value of this vev increases along the dashed line from zero (at the intersection point with the solid lines), up to large values. This means that  the phase transition at  $\gamma_{I}=\gamma_{B}\gtrsim 1$ can be  studied by a GL expansion. In the next section we develop this expansion, which leads to an interesting analogy with ultracold bosonic atoms and will help to clarify the nature of this phase transition.

\section{Analysis of the phase transitions in the $\mu_{B}-\mu_{I}$ plane}
\label{sec:phase}
In order to understand the mechanisms underlying  the phase transitions shown in Fig.\ref{fig:phase_diagram}, we perform a GL expansion of the potential. As we will see,  the second order transitions can be understood within the standard GL theory, however the phase transition along the dashed line is somehow unconventional in that it can be explained through a GL expansion up to quadratic order in the fields. This is in contrast to scalar theories with a single complex field (and its conjugate), in which it is necessary to expand the (effective) potential to the sixth order in order in the fields to observe a first order transition. 

\begin{figure*}[t!]
\centering
\includegraphics[width=0.31\textwidth]{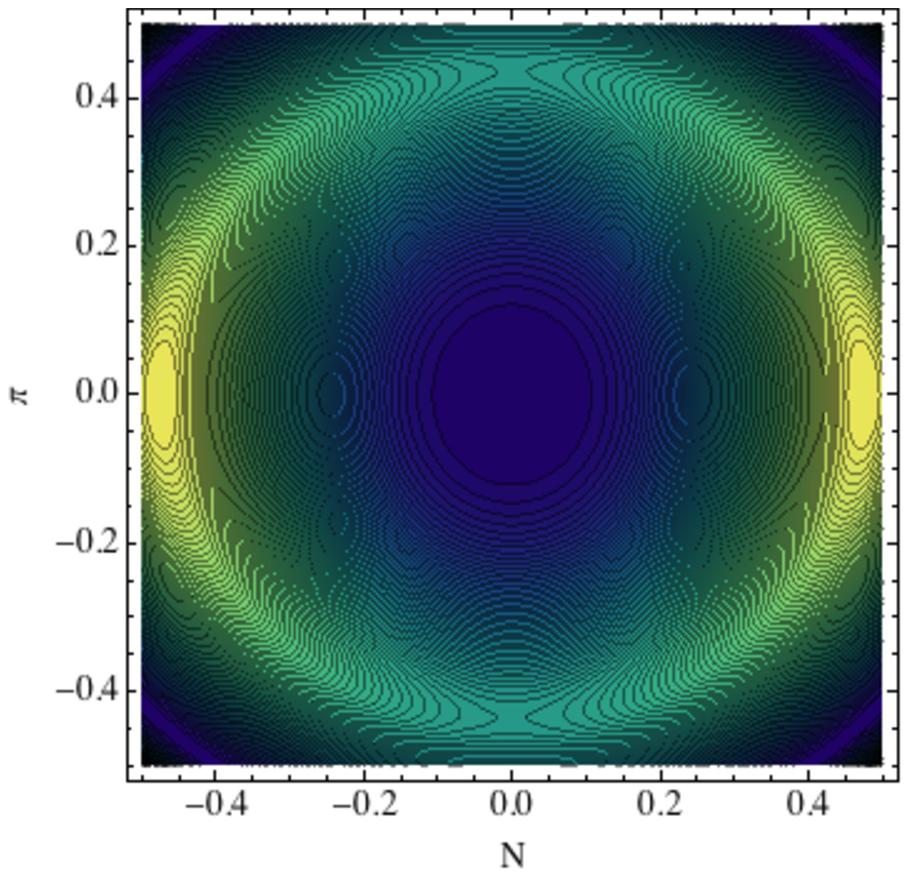}\,\,\,\,\,
\includegraphics[width=0.31\textwidth]{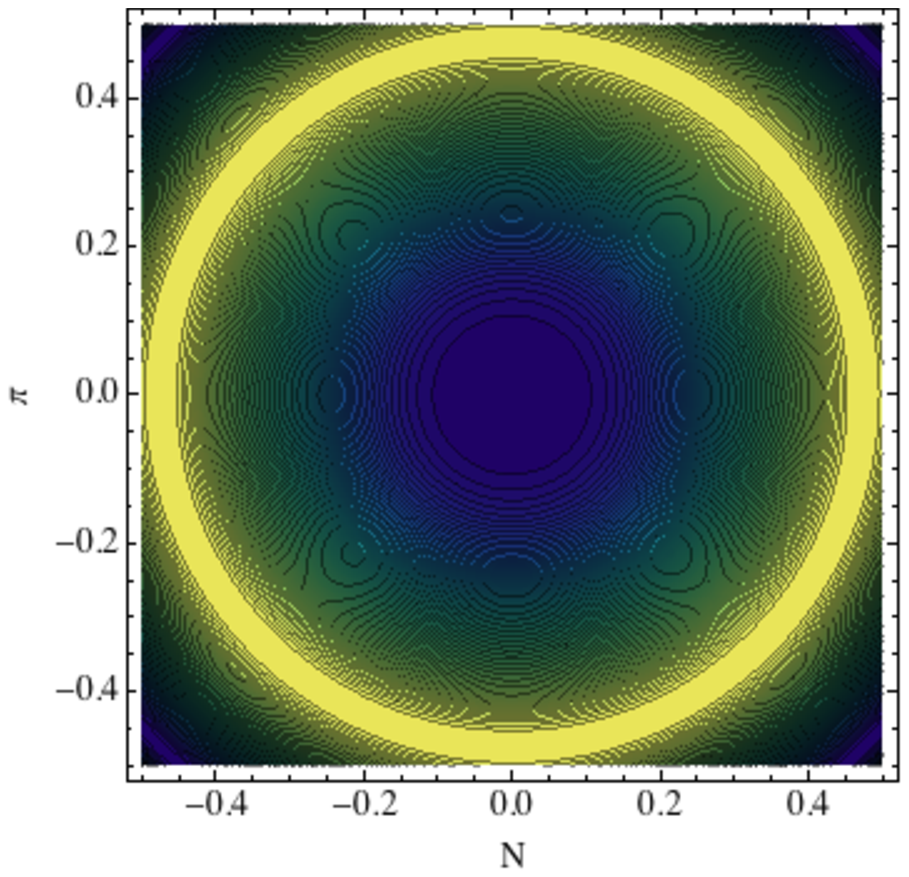}\,\,\,\,\,
\includegraphics[width=0.31\textwidth]{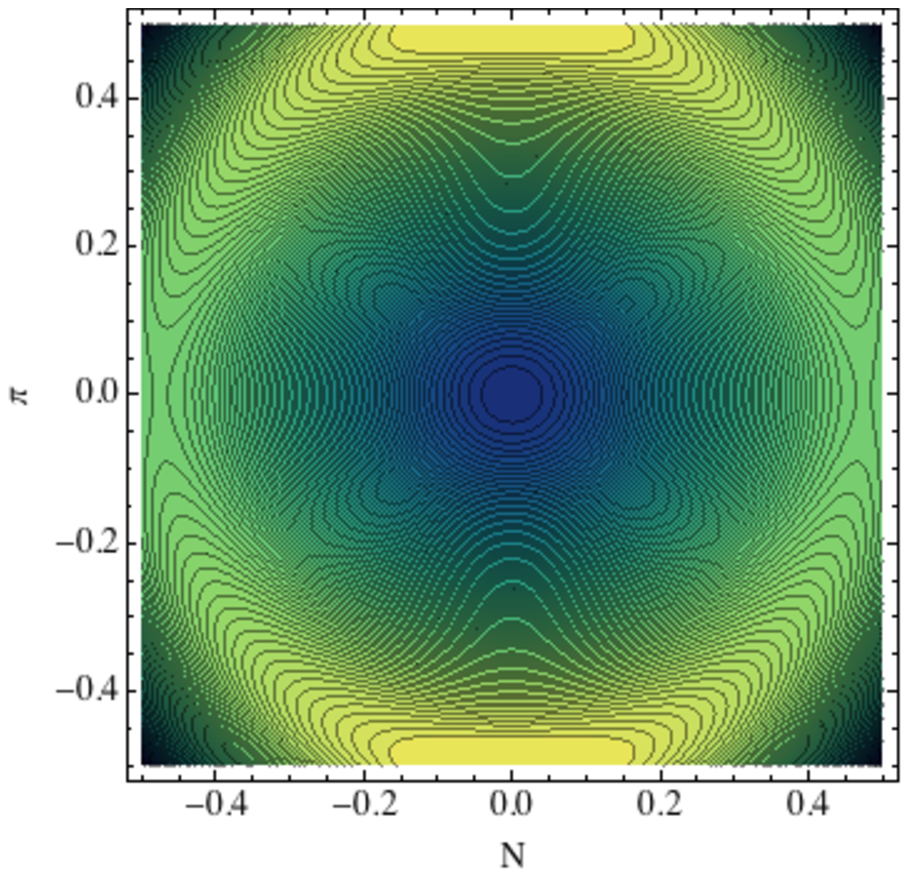}
\caption{Contour plots of the Ginzburg-Landau potential, see  Eq.~\eqref{eq:GL},  as a function of $N$ and $\pi$  for $\gamma_B=1.15$ and increasing values of  $\gamma_I$. The lighter (yellow) regions correspond to minima. Left: 
case with $\gamma_I=1.12$. The minimum is attained when the $N$ fields gets a nonvanishing vev and $\pi=0$.
 Center: case with $\gamma_I=\gamma_B=1.15$, the minimum has a degenerate $O(2)$ symmetry.  Right: case with $\gamma_I=1.18$. The $\pi$ condenses while $N=0$. }
\label{fig:contours}       
\end{figure*}

Upon expanding the potential in Eq.~\eqref{eq:potential} in the diquark and pion fields we obtain
\begin{align}
\frac{V}{f_{\pi}^{2}m_{\pi}^{2}}=&-\left (\gamma_{B}^{2}-1 \right ) |N|^{2}-\left (\gamma_{I}^{2}-1 \right ) |\pi|^{2}\nonumber \\
&+\frac{1}{3f_{\pi}^{2}}\left (-\frac{1}{2}+2\gamma_{B}^{2}\right )|N|^{4}+\frac{1}{3f_{\pi}^{2}}\left (-\frac{1}{2}+2\gamma_{I}^{2}\right )|\pi|^{4}\nonumber \\
&+\frac{2}{3f_{\pi}^{2}}\left (-\frac{1}{2}+\gamma_{B}^{2}+\gamma_{I}^{2}\right )|N|^{2}|\pi|^{2}+\dots\,,\label{eq:GL}
\end{align}
where  $N$ and $\pi$ are defined in Eqs.~\eqref{eq:N_pi}. The dots correspond to the neglected terms of order $|N|^n |\pi|^m$ with $n+m>4$. This expansion is  under control for small values of the diquark and pion condensates, meaning that it effectively describes the region $\gamma_B < 1 + \epsilon$ and   $\gamma_I < 1 + \epsilon $, with $0 <\epsilon \ll1$. Therefore, Eq.~\eqref{eq:GL} can be used to explore all the second order phase transition lines and a small region around the first order phase transition line with  $\gamma_B \gtrsim 1$ and $\gamma_I \gtrsim 1$.  In other words, this expansion is not valid along the whole dashed transition line in Fig.~\ref{fig:phase_diagram}, but only  for the region close to the point where the dashed line  touches the solid lines.

The expanded version shows, as expected, a second order phase transition at $\gamma_{B}=1$ or $\gamma_{I}=1$. 
When one of the coefficients of the quadratic term becomes negative  either the diquark or the pion field acquires a vacuum expectation value.  The corresponding values of the condensates agree with those reported in Eqs.~\eqref{eq:cond1} and~\eqref{eq:cond2} for  $\gamma_{B} \gtrsim 1$ and $\gamma_{I}\gtrsim 1$, respectively. This is the standard behavior for a GL expansion, which is expected to correctly reproduce second order phase transitions.

The  phase transition corresponding  to the case in which both  $\gamma_{B} \gtrsim 1$ and $\gamma_{I}\gtrsim 1$ is illustrated in Fig.~\ref{fig:contours}.  In this figure we keep fixed $\gamma_{B}$ and increase $\gamma_I$, from left to right, from values of $\gamma_I$ below $\gamma_{B}$, left panel, to $\gamma_I=\gamma_B$, central panel, to $\gamma_I>\gamma_B$, right panel. What happens is different from what one would expect in a standard GL theory describing a weak first order phase transition. A first order phase transition should be driven by a competition  between  two different condensates corresponding to local minima of the potential which are separated by a potential barrier. In this case the potential barrier vanishes at  $\gamma_I=\gamma_B$. The two condensates are clearly visible in Fig.~\ref{fig:contours}, indeed   when $\gamma_B > \gamma_{I}$, left panel, the minimum at $|N|\neq 0$  becomes the absolute minimum; conversely, when $\gamma_{I} > \gamma_{B}$, right panel, the minimum corresponding to  $|\pi| \neq 0$ becomes the absolute minimum. 
However, when $\gamma_{B}=\gamma_{I}=\gamma$, central panel, there is no energy barrier separating the two minima  along the bright (yellow) circle, indeed the potential becomes a function of $|N|^{2}+|\pi|^{2}$ but not of $|N|$ and $|\pi|$ separately:
\begin{equation}
\label{potentialequal}
\begin{split}
\left.\frac{V}{f_{\pi}^{2}m_{\pi}^{2}}\right |_{\gamma_B=\gamma_I=\gamma} =&-\left (\gamma^{2}-1 \right ) (|N|^{2}+|\pi|^{2})\\
&+\frac{1}{3f_{\pi}^{2}}\left (-\frac{1}{2}+2\gamma^{2}\right )(|N|^{2}+|\pi|^{2})^{2}+\cdots
\end{split}
\end{equation}
and this structure generalizes to all orders in the field expansion as is evident from Eq.~(\ref{eq:potential}), which becomes:
\begin{equation}
V=-\frac{f_{\pi}^{2}m_{\pi}^2}{2}\left [2(\cos\rho-1)+\gamma^{2}\sin^{2}\rho\right ] \ .
\end{equation}
From the expanded version of the potential in Eq.~(\ref{potentialequal}) it is clear that for  $\gamma_{I}=\gamma_{B}$, the vev is acquired by $|N|^{2}+|\pi|^{2}$ but not by $|N|$ and $|\pi|$ separately unlike in a standard GL theory with two species and a $U(1)\times U(1)$ symmetry~\cite{Adhikari:2017oxb, Forgacs:2016iva}.

\subsection{Two-Component Gross-Pitaevskii Lagrangian}
Since the GL theory describing the phase transitions in Fig.~\ref{fig:phase_diagram} is quartic in the fields,  close to the  phase transition lines it can mapped to a  Gross-Pitaevskii (GP) form. To this end it is useful to rescale the  fields as follows \begin{align}
 \psi_{1}=\sqrt{4 f_{\pi}^{2}\mu_B} N\,,\qquad
 \psi_{2}=\sqrt{4 f_{\pi}^{2}\mu_I} \pi^*\,.
 \end{align}
Along the $\gamma_B=1$ line,  working up to quadratic order in the fields and ignoring second derivatives in time, see the discussion in~\cite{Carignano:2016lxe}, we obtain  
\be
\mathcal{L}_{\rm GP}^B=i\psi_{1}^{*}\partial_{0}\psi_{1}+\mu_{{\rm eff,}1} |\psi_{1}|^{2}+\psi_{1}^{*}\frac{\nabla^{2}}{2M_{1}}\psi_{1}-\frac{g_{1}}{2}|\psi_{1}|^{4}\,,
\ee
where 
\be\label{eq:coeff1} g_1= \frac{4\mu_B^{2}-m_{\pi}^{2}}{24 f_{\pi}^{2}\mu_B^{2}}\,,\qquad M_1=\mu_B\,,\qquad  \mu_{{\rm eff,}1}=\frac{\mu_B^{2}-m_{\pi}^{2}}{2\mu_B}.\ee
Analogous expressions hold along the $\gamma_I=1$ line, where the corresponding GP Lagrangian is
\be
\mathcal{L}_{\rm GP}^B=i\psi_{2}^{*}\partial_{0}\psi_{2}+\mu_{{\rm eff,}2} |\psi_{2}|^{2}+\psi_{2}^{*}\frac{\nabla^{2}}{2M_{2}}\psi_{2}-\frac{g_{2}}{2}|\psi_{2}|^{4}\,,
\ee
with \be\label{eq:coeff2} g_2= \frac{4\mu_I^{2}-m_{\pi}^{2}}{24f_{\pi}^{2}\mu_I^{2}}\,,\qquad M_2=\mu_I\,,\qquad \mu_{{\rm eff,}2}=\frac{\mu_I^{2}-m_{\pi}^{2}}{2\mu_I}.\ee A relevant aspect of these   equations is that all the GP coefficients depend on the chemical potentials, meaning that approaching the phase transition all the properties of the system change. On the other hand, in condensed matter systems, one can typically only change  the effective chemical potential and in restricted cases the coupling constant.

The two second order phase transitions are analogous  to the case discussed in~\cite{Carignano:2016lxe} for the pion condensation in two-flavor, three color QCD. More interesting is the analysis close to  the dashed line of Fig.~\ref{fig:contours}, in the region where $\gamma_{I} \sim \gamma_{B}\gtrsim 1$.  In this case a two-species scenario is realized with a  corresponding  two-species effective GP Lagrangian 
\begin{align}
\mathcal{L}_{\rm GP}=&i\psi_{1}^{*}\partial_{0}\psi_{1}+\mu_{{\rm eff,}1} |\psi_{1}|^{2}+\psi_{1}^{*}\frac{\nabla^{2}}{2M_{1}}\psi_{1}\\
&+i\psi_{2}^{*}\partial_{0}\psi_{2}+\mu_{{\rm eff,}2} |\psi_{2}|^{2}+\psi_{2}^{*}\frac{\nabla^{2}}{2M_{2}}\psi_{2}\\
&-\frac{g_{1}}{2}|\psi_{1}|^{4}-\frac{g_{2}}{2}|\psi_{2}|^{4}-g_{12}|\psi_{1}|^{2}|\psi_{2}|^2\,,
\end{align}
where  
\begin{align}
g_{12}&=\frac{2(\mu_B^{2}+\mu_I^{2})-m_{\pi}^{2}}{24f_{\pi}^{2}\mu_B\mu_I}\,,
\end{align}
corresponds to the interaction between the two condensates;   all the other coefficients are reported in Eqs.~\eqref{eq:coeff1} and \eqref{eq:coeff2}.
The GP Lagrangian above has a stable ground state only if the interaction between the condensates is repulsive,  meaning that $g_{12}>0$, which is realized for $\mu_B^{2}+\mu_I^{2}>\frac{m_{\pi}^{2}}{2}$. Given that   $g_{1}g_{2}\le g_{12}^{2}$ the repulsion between the two condensates is stronger than the condensate self-interaction. In condensed matter systems this inequality implies  that the two-species, $\psi_{1}$ and $\psi_{2}$  segregate~\cite{Ho:1996zz, Ao:1998zz, Son:2001td}, see~\cite{Ao:1998zz} for a simple mean field derivation based on energetic considerations. However, segregation can only happen  assuming fixed condensation numbers.  In our case the number densities cannot be simultaneously nonzero, meaning that the two broken phases are effectively separated and cannot be simultaneously realized. We do not exclude that inhomogeneous phases might be realized, see for example~\cite{Andersen:2018osr}, but in the present study we focus on homogeneous phases.
In any case,  the nature of the ground state of the GP Lagrangian for $\mu_I=\mu_B =\mu >1$ is fundamentally different from the condensed matter analog. The reason is that  $|\psi_{1}|$ and $|\psi_{2}|$ are not separately the vevs  but the sum of their magnitudes squared gets a vev:
\begin{equation}
\langle |\psi_{1}|^{2}+|\psi_{2}|^{2}\rangle=\frac{\mu_{\rm eff}}{g}=\frac{12 f_{\pi}^{2}\mu(\mu^{2}-m_{\pi}^{2})}{4\mu^{2}-m_{\pi}^{2}}\ ,
\end{equation}
if $\mu>m_{\pi}$ and zero otherwise.
Along the dashed line  the order parameter  is an SU(2) doublet unlike the case with $\mu_{I}\neq\mu_{B}$ when the order parameter is a complex $U(1)$ field.
This is most easily seen by constructing an $SU(2)$ doublet using $\psi_{1}$ and $\psi_{2}$
\begin{equation}
\Psi=\begin{pmatrix}
  \psi_{1} \\
  \psi_{2}
 \end{pmatrix}\,, 
 \end{equation}
and rewriting the two-species Lagrangian in the following form:
\begin{equation}
\mathcal{L}=i\Psi^{\dagger}\partial_{0}\Psi-\frac{1}{2M}\nabla\Psi^{\dagger}\nabla\Psi+\mu_{\rm eff}\Psi^{\dagger}\Psi-\frac{g}{2}(\Psi^{\dagger}\Psi)^{2}\ .
\end{equation}
It is clear from the form of the Lagrangian that it possesses the following symmetry structure:
\begin{equation}
\begin{split}
SU(2)_{\rm global}:\ &\Psi\rightarrow e^{i\alpha_{a}\tau^{a}}\Psi\\
U(1)_{\rm global}:\ &\Psi\rightarrow e^{i\alpha}\Psi\ ,
\end{split}
\end{equation}
where $\tau^{a}$ with $a=1,2,3$ are the Pauli matrices.
The symmetry group of this two-species theory is larger than of the single species Gross-Pitaevskii Lagrangian, which only possesses a $U(1)_{\rm global}$ symmetry if ungauged and a $U(1)_{\rm local}$ symmetry if gauged. For completeness, the equation of motion in terms of the field $\Psi$ is given by
 \begin{equation}
 i\partial_{0}\Psi=-\frac{1}{2M}\nabla^{2}\Psi-\mu_{\rm eff}\Psi+g(\Psi^{\dagger}\Psi)\Psi\ .
 \end{equation}
\subsection{Full Lagrangian}
We now turn to the full Lagrangian of the theory  in terms of the radial degree of freedom $\rho$ and the $\hat{\phi}_{i}$:
\begin{equation}
\begin{split}
\mathcal{L}&=\frac{f_\pi^2}{2}\left [\partial_{\mu}\rho\partial^{\mu}\rho+\sin^{2}\rho\partial_{\mu}\hat{\phi}_{i}\partial^{\mu}\hat{\phi}_{i}-2m_{\pi}\sin^{2}\rho\right.\\
&\left.\left (\gamma_{B}(\hat{\phi}_{1} \partial_0 \hat{\phi}_{3} - \hat{\phi}_{3} \partial_0 \hat{\phi}_{1})+\gamma_{I}(\hat{\phi}_2 \partial_0 \hat{\phi}_4 - \hat{\phi}_4 \partial_0 \hat{\phi}_2)\right )\right ]\\
&+f_{\pi}^{2} m_\pi^2 (\cos\rho-1)+\frac{f_\pi^2m_{\pi}^{2}}{2} \sin^{2}\rho\left [ \gamma_{B}^2 (\hat{\phi}_1^2 + \hat{\phi}_{3}^{2})\right.\\
&\left.+\gamma_{I}^{2} (\hat{\phi}_{2}^2 + \hat{\phi}_{4}^2)\right ]\ .\label{eq:full_lagr}
\end{split}
\end{equation}
For simplicity we have set $\hat{\phi}_{5}=0$, since the neutral pion does not condense in any of the ground states. 
It is instructive to express  $\hat{\phi}_{i}$ in terms of angular variables (Hopf coordinates):
\begin{align}
\hat{\phi}_{1}=&\cos\theta\cos\alpha_{N},\ \hat{\phi}_{3}=\cos\theta\sin\alpha_{N}\,,\\
\hat{\phi}_{2}=&\sin\theta\cos\alpha_{\pi},\ \hat{\phi}_{4}=\sin\theta\sin\alpha_{\pi}\ .
\end{align}
Note the $\alpha_{\pi}$ and  $\alpha_{N}$ are the phases of the pion and diquark fields, respectively, and that $\theta$ determines the relative size of the pion  and diquark  fields. In other words, $\theta$ is the angular coordinate in the $\pi ,N$ plane of Fig.~\ref{fig:contours}. 

\begin{figure}
  \includegraphics[width=0.45\textwidth]{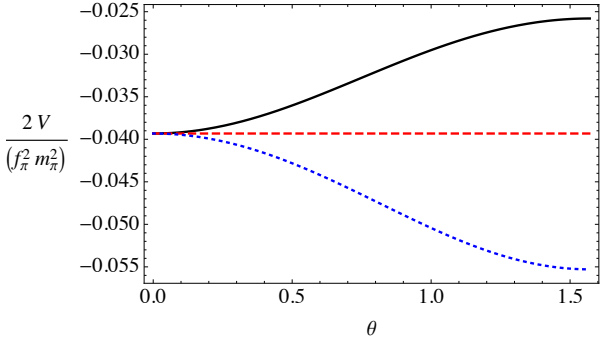}
\caption{Potentials as a function of the angular coordinate for the same valus of the chemical potentials used in Fig.~\ref{fig:contours}, that is $\gamma_B=1.15$ and three different values of $\gamma_I$. The solid black line corresponds to $\gamma_I=1.12$, the dashed red line to   $\gamma_I=1.15$ and the dotted blue line to  $\gamma_I=1.18$.}
\label{fig:potential_theta}       
\end{figure}

We will first consider unequal values of $\gamma_B$ and $\gamma_I$ and then let these two quatities be equal.
The angular stationary points of the potential are  given by
\be
\sin 2 \theta=0\,,
\ee 
while the radial coordinate has stationary points
\begin{align}
\sin(\rho)&=0\,, \\
\label{eq:rhovev}\cos(\rho)&=\frac{1}{\gamma_B^2 \cos^2\theta+\gamma_I^2 \sin^2\theta}\,. 
\end{align}
Restricting to the first quadrant, we have that the point in $\bar\rho=0$ is a maximum. The  points with $\theta=0$ and $\theta=\pi/2$ can be a minimum or a saddle point. If $\gamma_B > \gamma_I$, the minimum is at $\theta=0$, and the saddle point at $\theta=\pi/2$.  The opposite happens for $\gamma_B < \gamma_I$.
The potential term is 
\begin{equation}
\label{eq:VVV}
V=-\frac{f_{\pi}^{2}m_{\pi}^2}{2}\left [2(\cos\rho-1)+\sin^{2}\rho (\gamma_B^2 \cos^2\theta + \gamma_I^2 \sin^2\theta)\right ] \ .
\end{equation}
Upon substituting in the above expression the value of the vev of the radial coordinate given in Eq.~\eqref{eq:rhovev}, we obtain the results reported in Fig.~\ref{fig:potential_theta}, for the same values of the chemical potential used in Fig.~\ref{fig:contours}. The three plots in  Fig.~\ref{fig:potential_theta} clearly show that no potential barrier is present along the theta direction and that the two minima in $\theta=0$ and in $\theta=\pi/2$ do not coexist. 

We can now consider the fluctuations around the minimum, writing $\rho=\bar\rho + \chi$ and $\theta=\bar\theta +h$, then upon substituting these expression in  Eq.~\eqref{eq:full_lagr} we find that the masses of these fields  are given by 
\begin{align}
m_h^2&= m_\pi^2 |\gamma_B^2-\gamma_I^2| \sin^2\bar\rho\,,\\
m_\chi^2 & = m_\pi^2 \frac{\sin^2\bar\rho}{\cos\bar\rho}\label{eq:mchi}\,,
\end{align}
where the mass of the $\chi$ field agrees with the result reported in \cite{Carignano:2016lxe}, while the mass of the $h$ field is a feature of the present model. Note that $m_h < m_\chi$ and that $m_h$ vanishes along the dashed line of Fig.~\ref{fig:phase_diagram}. Since the $\chi$ mode is the heaviest one, it can be integrated out. We will explicitly do this in the following section for the peculiar $\mu_B=\mu_I$ case.

\section{Effective Lagrangian at $\mu_{B}=\mu_{I}$}
\label{sec:effective}
In the case with $\gamma_B\neq \gamma_I$, the Lagrangian possesses a $U(1)\times U(1)$ symmetry~\cite{Son:2001td}. However, for the special case of equal baryon and isospin chemical potentials, i.e. $\mu_{B}=\mu_{I}$, the potential energy is independent of the angular variable $\theta$, see Eq.~\eqref{eq:VVV}. For any small perturbation of the difference between the two chemical potentials, the system will spontaneously choose a direction corresponding to an angle $\bar\theta$, which is either $\overline{\theta}=0$ or $\overline{\theta}=\frac{\pi}{2}$. For an illustration of the degeneracy and the consequent symmetry breaking, see the center plot in Fig.~\ref{fig:contours} and the plots to the left and right, respectively. In the special case with equal chemical potentials, the Lagrangian has an expanded $SU(2)_{\rm global}$ symmetry and the effective Lagrangian can be written in terms of the $SU(2)$ doublet field $\Phi\equiv\begin{pmatrix}\pi\\N\end{pmatrix}$ and its conjugate transpose. In other words, the pion and diquark field can be transformed into each other through a $SU(2)_{\rm global}$ as  previously argued by the GL analysis. Furthermore, the heavy radial degree of freedom, $\chi$, can be integrated out, which results in an effective Lagrangian with three (angular) degrees of freedom~\cite{Schafer:2001bq, Miransky:2001tw} 
\begin{align}
\mathcal{L}_{\rm eff}=&\frac{f_{\pi}}{2}\sin^{2}\overline{\rho}\left [\frac{1}{2}\left \{(\partial_{0}\theta)^{2}-(\partial_{i}\theta)^{2} \right \}\right.\nonumber\\
&\left.+\frac{\cos^{2}\theta}{2}\left \{\frac{\gamma^{4}-1+4\cos^{2}\theta}{\gamma^{4}-1}(\partial_{0}\alpha_{N})^{2}-(\partial_{i}\alpha_{N})^{2}\right \}\right.\nonumber\\
&\left.+\frac{\sin^{2}\theta}{2}\left \{\frac{\gamma^{4}-1+4\sin^{2}\theta}{\gamma^{4}-1}(\partial_{0}\alpha_{\pi})^{2}-(\partial_{i}\alpha_{\pi})^{2}\right \}\right.\nonumber\\
&\left.-\frac{\sin^{2}2\theta}{\gamma^{4}-1}\partial_{0}\alpha_{N}\partial_{0}\alpha_{\pi}\right ]\,,
\end{align}
which is valid for energy scales below the $\chi$ mass in Eq.~\eqref{eq:mchi}, and accounts for all the propagating and interaction terms of the soft modes. Similar Lagrangians have been derived in condensed matter systems~\cite{PhysRevLett.89.170403, Ho:1996zz, PhysRevLett.81.5257, PhysRevLett.85.5488, 1998JPSJ...67.1822O}, see \cite{Kawaguchi:2012ii} for a review. However, there is an importance difference between this Lagrangian and those in condensed matter systems. The $\theta$ degree of freedom (in condensed matter systems) picks up a vev, thereby allowing the identification of sound modes, which is a linear combination of $\alpha_{N}$ and $\alpha_{\pi}$. We are not aware of any such mechanism in the context of two-color $\chi$PT. If such a mechanism exists, expanding around the vev, i.e. $\theta=\overline{\theta}+h$, one could identify two (soft) modes of sound: a fast mode, $\alpha_{-}=\alpha_{N}-\alpha_{\pi}$, which propagates at the speed of light and a slower mode $\alpha_{+}=\sin^{2}\overline{\theta}\alpha_{N}+\cos^{2}\overline{\theta}\alpha_{\pi}$, which propagates at $c_{s}=\sqrt{\frac{\gamma^{4}-1}{\gamma^{4}+3}}$. (Note that when $\mu_{B}\neq\mu_{I}$, i.e. $\overline{\theta}=0$ or $\overline{\theta}=\frac{\pi}{2}$, we can identify $\alpha_{N}$ and $\alpha_{\pi}$ as the modes propagating at $c_{s}$, which is as expected from three-color two-flavor $\chi$PT~\cite{Carignano:2016lxe}.) However, thermodynamic analysis of two-color  $\chi$PT (for details, please refer to \ref{AppendixA}) leads to the following differential relation between pressure $p$ and the energy density $\epsilon$ even for equal chemical potentials $\mu_{B}=\mu_{I}$,
\begin{equation}
\sqrt{\frac{\partial p}{\partial \epsilon}}=\sqrt{\frac{\gamma^{4}-1}{\gamma^{4}+3}}\ .
\end{equation}
While this relation can be interpreted as the speed of sound for the condensed pion or diquark phase when ($\mu_{B}\neq \mu_{I}$), such an interpretation, as far as we are aware, is only possible for $\mu_{B}=\mu_{I}$ if $\theta$ possessed a vev.

\section{Conclusions}
We have shown that two-color, two-flavor chiral perturbation theory possesses a global $U(1)\times SU(2)$ symmetry when isospin chemical potential equals the diquark chemical potential. The phase transition from the superfluid diquark phase at ($\mu_{B}>\mu_{I}$) to the superfluid pion at ($\mu_{I}>\mu_{B}$) involves an intermediary phase (when $\mu_{B}=\mu_{I}$), where the vev of the quantity $\pi^*\pi+N^*N$ is fixed but that of the pion and diquark condensates are not. This results in a peculiar phase transition, in which the number densities are discontinuous, as in a first order phase transition, but there exists a flat direction of the potential, as in second order phase transitions. Indeed, one of the two minima turns into a saddle point along the dashed phase transition line of Fig.~\ref{fig:phase_diagram}. Finally, we constructed an effective theory in terms of the $SU(2)$ parameters by integrating out the radial modes. We expect similar results to  apply to three-color, three-flavor chiral perturbation theory when both isospin and strange chemical potentials are varied~\cite{Kogut:2001id}. 

There are a number of interesting open problems that deserve further attention, including the study of 
inhomogeneous phases, possibly using the extension of the GL Lagrangian of~\cite{Carignano:2017meb},  semi-local strings~\cite{Achucarro:1999it,Chernodub:2010sg} in external magnetic fields, which we plan to pursue in future work. 

\appendix
\section{Thermodynamics}
\label{AppendixA}
The ground state satisfies $\rho=\overline{\rho}$ for each of the three condensed phases resulting in a potential that is a function of $\sum_{i} \hat{\phi}_{i}^{2}$. Since this sum adds to one, the thermodynamic quantities have the same values in each of the condensed phases. As such, we will use a generic variable $\mu$ for the chemical potential and $\gamma\equiv\frac{\mu}{m_{\pi}}$. The $\mu$ should be taken to mean the appropriate chemical potential: $\mu=\mu_{I}=\mu_{B}$, $\mu=\mu_{I}>\mu_{B}$ or $\mu=\mu_{B}>\mu_{I}$. The resulting size of the potential in terms of $\gamma$ in the vacuum state is:
\begin{equation}
V(\overline{\rho})=V(\gamma)=-\frac{f_{\pi}^{2}m_{\pi}^{2}}{2}\left (\frac{\gamma^{2}-1}{\gamma} \right )^{2}\ ,
\end{equation}
which is valid when $\gamma\ge1$. The potential has been normalized to be zero in the normal vacuum. The resulting energy density is 
\begin{equation}
\epsilon(\gamma)=V(\overline{\rho})+\mu n(\gamma)=\frac{f_{\pi}^{2}m_{\pi}^{2}}{2}\left (\frac{\gamma^{4}+2\gamma^{2}-3}{\gamma^{2}} \right )\ ,
\end{equation}
which can be written solely in terms of $\gamma$ and the number density as:
\begin{equation}
\epsilon(n)=-m_{\pi}^{2}f_{\pi}^{2}\left (\sqrt{1-\frac{n(\gamma)}{m_{\pi}f_{\pi}^{2}\gamma}}-1 \right )+\gamma\frac{m_{\pi}n(\gamma)}{2}\ ,
\end{equation}
with the number density in each of the condensed phases as follows:
\begin{equation}
n(\gamma)=-\frac{1}{m_{\pi}}\frac{\partial V}{\partial \gamma}=f_{\pi}^{2}m_{\pi}\gamma\left (1-\frac{1}{\gamma^{4}} \right )\ .
\end{equation}
The resulting pressure in the normal and condensed phases is
\begin{equation}
\begin{split}
p&=n\frac{\partial \epsilon(n)}{\partial n}-\epsilon(n)\\
&=-\frac{m_{\pi}\left [n(\gamma)+2m_{\pi}f_{\pi}^{2}\gamma\left (\sqrt{1-\frac{n(\gamma)}{m_{\pi}f_{\pi}^{2}\gamma}}-1\right ) \right ]}{2\gamma\sqrt{1-\frac{n(\gamma)}{m_{\pi}f_{\pi}^{2}\gamma}}}\ ,
\end{split}
\end{equation}
which can be used to compute the speed of sound in both the pion and diquark condensed phases (including the $\mu_{I}=\mu_{B}$ line)
\begin{equation}
\label{eq:speedofsound}
c_{s}=\sqrt{\frac{\partial p}{\partial\epsilon}}=\sqrt{\frac{\gamma^{4}-1}{\gamma^{4}+3}}=\sqrt{\frac{\mu^{4}-m_{\pi}^{4}}{\mu^{4}+3m_{\pi}^{4}}}\,.
\end{equation}
and is identical to that of the pion condensed phase in three-color QCD~\cite{Son:2000xc}.

\section*{Acknowledgements}
P.A. and S.B.B. would like to acknowledge the research support provided through the Collaborative Undergraduate Research and Inquiry (CURI, 2017) program at St. Olaf College, where this work was initiated. P.A. would like to acknowledge Stefano Carignano and Charles Cunningham for some useful comments on related physics. Finally, P.A. would like to thank undergraduate students Emma Dawson, Jaehong Choi, Benjamin Wollant and Colin Scheibner for many useful discussions. \\


\begin{thebibliography}{100}
\providecommand{\url}[1]{{#1}}
\providecommand{\urlprefix}{URL }
\expandafter\ifx\csname urlstyle\endcsname\relax
  \providecommand{\doi}[1]{DOI \discretionary{}{}{}#1}\else
  \providecommand{\doi}{DOI \discretionary{}{}{}\begingroup
  \urlstyle{rm}\Url}\fi

\bibitem{Barbour:1986jf}
I.~Barbour, N.E. Behilil, E.~Dagotto, F.~Karsch, A.~Moreo, et~al., Nucl.Phys.
  \textbf{B275}, 296 (1986).
\newblock \doi{10.1016/0550-3213(86)90601-2}

\bibitem{Barbour:1997ej}
I.M. Barbour, S.E. Morrison, E.G. Klepfish, J.B. Kogut, M.P. Lombardo, Nucl.
  Phys. Proc. Suppl. \textbf{60A}, 220 (1998).
\newblock \doi{10.1016/S0920-5632(97)00484-2}

\bibitem{Alford:1998sd}
M.G. Alford, A.~Kapustin, F.~Wilczek, Phys.Rev. \textbf{D59}, 054502 (1999).
\newblock \doi{10.1103/PhysRevD.59.054502}

\bibitem{Kogut:2002zg}
J.~Kogut, D.~Sinclair, Phys.Rev. \textbf{D66}, 034505 (2002).
\newblock \doi{10.1103/PhysRevD.66.034505}

\bibitem{Detmold:2012wc}
W.~Detmold, K.~Orginos, Z.~Shi, Phys. Rev. \textbf{D86}, 054507 (2012).
\newblock \doi{10.1103/PhysRevD.86.054507}

\bibitem{Detmold:2008yn}
W.~Detmold, K.~Orginos, M.J. Savage, A.~Walker-Loud, Phys. Rev. \textbf{D78},
  054514 (2008).
\newblock \doi{10.1103/PhysRevD.78.054514}

\bibitem{Cea:2012ev}
P.~Cea, L.~Cosmai, M.~D'Elia, A.~Papa, F.~Sanfilippo, Phys. Rev. \textbf{D85},
  094512 (2012).
\newblock \doi{10.1103/PhysRevD.85.094512}

\bibitem{Endrodi:2014lja}
G.~Endr{\"o}di, Phys. Rev. \textbf{D90}(9), 094501 (2014).
\newblock \doi{10.1103/PhysRevD.90.094501}

\bibitem{Brandt:2017oyy}
B.B. Brandt, G.~Endrodi, S.~Schmalzbauer,  arXiv:1712.08190v1 [hep-lat] (2017).

\bibitem{Adhikari:2015wva}
P.~Adhikari, T.D. Cohen, J.~Sakowitz, Phys. Rev. \textbf{C91}(4), 045202
  (2015).
\newblock \doi{10.1103/PhysRevC.91.045202}

\bibitem{Manohar:1998xv}
A.V. Manohar, in \emph{{Probing the standard model of particle interactions.
  Proceedings, Summer School in Theoretical Physics, NATO Advanced Study
  Institute, 68th session, Les Houches, France, July 28-September 5, 1997. Pt.
  1, 2}} (1998), pp. 1091--1169

\bibitem{Kogut:1999iv}
J.~Kogut, M.A. Stephanov, D.~Toublan, Phys.Lett. \textbf{B464}, 183 (1999).
\newblock \doi{10.1016/S0370-2693(99)00971-5}

\bibitem{Kogut:2000ek}
J.~Kogut, M.A. Stephanov, D.~Toublan, J.~Verbaarschot, A.~Zhitnitsky,
  Nucl.Phys. \textbf{B582}, 477 (2000).
\newblock \doi{10.1016/S0550-3213(00)00242-X}

\bibitem{Kogut:2001na}
J.B. Kogut, D.K. Sinclair, S.J. Hands, S.E. Morrison, Phys. Rev. \textbf{D64},
  094505 (2001).
\newblock \doi{10.1103/PhysRevD.64.094505}

\bibitem{Hands:2000ei}
S.~Hands, I.~Montvay, S.~Morrison, M.~Oevers, L.~Scorzato, J.~Skullerud, Eur.
  Phys. J. \textbf{C17}, 285 (2000).
\newblock \doi{10.1007/s100520000477}

\bibitem{Ratti:2004ra}
C.~Ratti, W.~Weise, Phys. Rev. \textbf{D70}, 054013 (2004).
\newblock \doi{10.1103/PhysRevD.70.054013}

\bibitem{Andersen:2010vu}
J.O. Andersen, T.~Brauner, Phys. Rev. \textbf{D81}, 096004 (2010).
\newblock \doi{10.1103/PhysRevD.81.096004}

\bibitem{Andersen:2014xxa}
J.O. Andersen, W.R. Naylor, A.~Tranberg, Rev. Mod. Phys. \textbf{88}, 025001
  (2016).
\newblock \doi{10.1103/RevModPhys.88.025001}

\bibitem{Adhikari:2016eef}
P.~Adhikari, J.O. Andersen, P.~Kneschke, Phys. Rev. \textbf{D95}(3), 036017
  (2017).
\newblock \doi{10.1103/PhysRevD.95.036017}

\bibitem{Buballa:2014tba}
M.~Buballa, S.~Carignano, Prog. Part. Nucl. Phys. \textbf{81}, 39 (2015).
\newblock \doi{10.1016/j.ppnp.2014.11.001}

\bibitem{Weinberg:1978kz}
S.~Weinberg, Physica \textbf{A96}, 327 (1979)

\bibitem{Pich:1998xt}
A.~Pich, in \emph{{Probing the standard model of particle interactions.
  Proceedings, Summer School in Theoretical Physics, NATO Advanced Study
  Institute, 68th session, Les Houches, France, July 28-September 5, 1997. Pt.
  1, 2}} (1998), pp. 949--1049

\bibitem{Holstein:2000ap}
B.R. Holstein, Nucl. Phys. \textbf{A689}, 135 (2001).
\newblock \doi{10.1016/S0375-9474(01)00828-4}

\bibitem{Scherer:2005ri}
S.~Scherer, M.R. Schindler,   \emph{A Primer for Chiral Perturbation Theory}, Springer-Verlag, Berlin, Heidelberg, (2012).

\bibitem{Son:2000xc}
D.~Son, M.A. Stephanov, Phys.Rev.Lett. \textbf{86}, 592 (2001).
\newblock \doi{10.1103/PhysRevLett.86.592}

\bibitem{Kogut:2001id}
J.~Kogut, D.~Toublan, Phys.Rev. \textbf{D64}, 034007 (2001).
\newblock \doi{10.1103/PhysRevD.64.034007}

\bibitem{Carignano:2016lxe}
S.~Carignano, L.~Lepori, A.~Mammarella, M.~Mannarelli, G.~Pagliaroli, Eur.
  Phys. J. \textbf{A53}(2), 35 (2017).
\newblock \doi{10.1140/epja/i2017-12221-x}

\bibitem{Loewe:2002tw}
M.~Loewe, C.~Villavicencio, Phys. Rev. \textbf{D67}, 074034 (2003).
\newblock \doi{10.1103/PhysRevD.67.074034}

\bibitem{Loewe:2004mu}
M.~Loewe, C.~Villavicencio, Phys.Rev. \textbf{D70}, 074005 (2004).
\newblock \doi{10.1103/PhysRevD.70.074005}

\bibitem{Loewe:2016wsk}
M.~Loewe, A.~Raya, C.~Villavicencio,   (2016)

\bibitem{Splittorff:2000mm}
K.~Splittorff, D.T. Son, M.A. Stephanov, Phys. Rev. \textbf{D64}, 016003
  (2001).
\newblock \doi{10.1103/PhysRevD.64.016003}

\bibitem{Smilga:1994tb} 
  A.~V.~Smilga and J.J.~M.~Verbaarschot,
  Phys. Rev.   \textbf{D51}, 829 (1995).
\newblock \doi{10.1103/PhysRevD.51.829}

\bibitem{Brauner:2006dv}
T.~Brauner, Mod. Phys. Lett. \textbf{A21}, 559 (2006).
\newblock \doi{10.1142/S0217732306019657}

\bibitem{Gasser:1983yg}
J.~Gasser, H.~Leutwyler, Annals Phys. \textbf{158}, 142 (1984).
\newblock \doi{10.1016/0003-4916(84)90242-2}

\bibitem{Adhikari:2017oxb}
P.~Adhikari, J.~Choi, Acta Phys. Polon. \textbf{B48}, 145 (2017).
\newblock \doi{10.5506/APhysPolB.48.145}

\bibitem{Forgacs:2016iva}
P.~Forg{\'a}cs, {\'A}.~Luk{\'a}cs, Phys. Rev. \textbf{D94}(12), 125018 (2016).
\newblock \doi{10.1103/PhysRevD.94.125018}

\bibitem{Ho:1996zz}
T.L. Ho, V.B. Shenoy, Phys. Rev. Lett. \textbf{77}, 3276 (1996).
\newblock \doi{10.1103/PhysRevLett.77.3276}

\bibitem{Ao:1998zz}
P.~Ao, S.T. Chui, Phys. Rev. \textbf{A58}, 4836 (1998).
\newblock \doi{10.1103/PhysRevA.58.4836}

\bibitem{Son:2001td}
D.T. Son, M.A. Stephanov, Phys. Rev. \textbf{A65}, 063621 (2002).
\newblock \doi{10.1103/PhysRevA.65.063621}

\bibitem{Andersen:2018osr}
J.O. Andersen, P.~Kneschke, arXiv:1802.01832v1 [hep-ph] (2018).

\bibitem{Schafer:2001bq}
T.~Sch{\"a}fer, D.T. Son, M.A. Stephanov, D.~Toublan, J.J.M. Verbaarschot,
  Phys. Lett. \textbf{B522}, 67 (2001).
\newblock \doi{10.1016/S0370-2693(01)01265-5}

\bibitem{Miransky:2001tw}
V.A. Miransky, I.A. Shovkovy, Phys. Rev. Lett. \textbf{88}, 111601 (2002).
\newblock \doi{10.1103/PhysRevLett.88.111601}

\bibitem{PhysRevLett.89.170403}
A.B. Kuklov, B.V. Svistunov, Phys. Rev. Lett. \textbf{89}, 170403 (2002).
\newblock \doi{10.1103/PhysRevLett.89.170403}.
\newblock
  \urlprefix\url{https://link.aps.org/doi/10.1103/PhysRevLett.89.170403}

\bibitem{PhysRevLett.81.5257}
C.K. Law, H.~Pu, N.P. Bigelow, Phys. Rev. Lett. \textbf{81}, 5257 (1998).
\newblock \doi{10.1103/PhysRevLett.81.5257}.
\newblock \urlprefix\url{https://link.aps.org/doi/10.1103/PhysRevLett.81.5257}

\bibitem{PhysRevLett.85.5488}
A.B. Kuklov, J.L. Birman, Phys. Rev. Lett. \textbf{85}, 5488 (2000).
\newblock \doi{10.1103/PhysRevLett.85.5488}.
\newblock \urlprefix\url{https://link.aps.org/doi/10.1103/PhysRevLett.85.5488}

\bibitem{1998JPSJ...67.1822O}
T.~{Ohmi}, K.~{Machida}, Journal of the Physical Society of Japan \textbf{67},
  1822 (1998).
\newblock \doi{10.1143/JPSJ.67.1822}

\bibitem{Kawaguchi:2012ii}
Y.~Kawaguchi, M.~Ueda, Phys. Rept. \textbf{520}, 253 (2012).
\newblock \doi{10.1016/j.physrep.2012.07.005}

\bibitem{Carignano:2017meb}
S.~Carignano, M.~Mannarelli, F.~Anzuini, O.~Benhar, Phys. Rev. \textbf{D97}(3),
  036009 (2018).
\newblock \doi{10.1103/PhysRevD.97.036009}

\bibitem{Achucarro:1999it}
A.~Achucarro, T.~Vachaspati, Phys. Rept. \textbf{327}, 347 (2000).
\newblock \doi{10.1016/S0370-1573(99)00103-9}.
\newblock [Phys. Rept.327,427(2000)]

\bibitem{Chernodub:2010sg}
M.N. Chernodub, A.S. Nedelin, Phys. Rev. \textbf{D81}, 125022 (2010).
\newblock \doi{10.1103/PhysRevD.81.125022}

\end{thebibliography}
\end{document}